\documentclass[12pt]{article}
\usepackage{times}
\usepackage{geometry}
\usepackage[utf8]{inputenc}
\usepackage{enumitem,amssymb}
\usepackage{ragged2e}

\usepackage{graphicx}
\usepackage{caption}
\usepackage{subcaption}
\usepackage{hyperref}
\usepackage[usenames]{color}
\usepackage{natbib}
\usepackage{aas_macros}
\newlist{thematic}{itemize}{8}
\setlist[thematic]{label=$\square$}
\usepackage{pifont}
\newcommand{\cmark}{\ding{51}}%
\newcommand{\done}{\rlap{$\square$}{\raisebox{2pt}{\large\hspace{1pt}\cmark}}%
\hspace{-2.5pt}}

\newcommand{\ANTARES}{{\textsc {antares}}}

\newcommand{\rob}[1]{\textcolor{black}{#1}}

\begin{document}
\raggedright
\huge
Astro2020 Activities, Projects, or State of the Profession Consideration White Paper \linebreak
The Data Lab: A Science Platform for the analysis of ground-based astronomical survey data
 \linebreak
\normalsize

\noindent \textbf{Thematic Areas:} \hspace*{30pt} \done Ground Based Project \hspace*{10pt} $\square$ Space Based Project \hspace*{20pt}\linebreak
\done Infrastructure Activity \hspace*{10pt} \done Technological Development Activity \linebreak
  $\square$  State of the Profession Consideration  \hspace*{10pt} $\square$ Other \linebreak
  
\textbf{Principal Author:}

Name: Knut A.G. Olsen	
 \linebreak						
Institution: National Optical Astronomy Observatory 
 \linebreak
Email: kolsen@noao.edu
 \linebreak
Phone: 520-318-8555 
 \linebreak
 
\textbf{Co-authors:} Adam Bolton (NOAO, abolton@noao.edu), St{\'e}phanie Juneau (NOAO, sjuneau@noao.edu), Robert Nikutta (NOAO, nikutta@noao.edu), Dara Norman (NOAO, dnorman@noao.edu), David Nidever (NOAO and Montana State  University, dnidever@noao.edu), Stephen Ridgway (NOAO, sridgway@noao.edu), Adam Scott (NOAO and Aerotek, ascott@noao.edu), and Benjamin Weaver (NOAO, baweaver@lbl.gov) for the NOAO Data Lab Team
  \linebreak

\textbf{Abstract  (optional):}
The next decade will feature a growing number of massive ground-based photometric, spectroscopic, and time-domain surveys, including those produced by DECam, DESI, and LSST.  The NOAO Data Lab was launched in 2017 to enable efficient exploration and analysis of large surveys, with particular focus on the petabyte-scale holdings of the NOAO Archive and their associated catalogs.  The Data Lab mission and future development align well with two of the NSF's Big Ideas,
namely {\it Harnessing Data for 21st Century Science and Engineering} and as part of a network to contribute to {\it Windows on the Universe: The Era of Multi-messenger Astrophysics}.  Along with other Science Platforms, the Data Lab will play a key role in scientific discoveries from surveys in the next decade, and will be crucial to maintaining a level playing field as datasets grow in size and complexity.


\pagebreak
\section{Why do we need the NOAO Data Lab?}
The 21st century
has seen massive growth in the popularity and impact of large-scale surveys in astronomy. Wide sky coverage and large sample sizes have become keys to addressing critical questions in cosmology, tracing galaxy evolution over the history of the Universe, probing the structure of the Milky Way and nearby galaxies, and using time series observations to discover 
variable or moving objects and transient events ranging from small bodies in our own solar system to stellar explosions in distant galaxies. Following in the footsteps of the Sloan Digital Sky Survey, imaging and spectroscopic surveys and survey instruments have proliferated, a trend driven by the needs for wide-area coverage and high statistical significance, which will continue as LSST, the top priority from the 2010 Decadal Survey, begins operations in the coming decade.



\begin{figure}
\centering
    \begin{subfigure}[b]{0.45\textwidth}
        \centering
        \includegraphics[width=\textwidth]{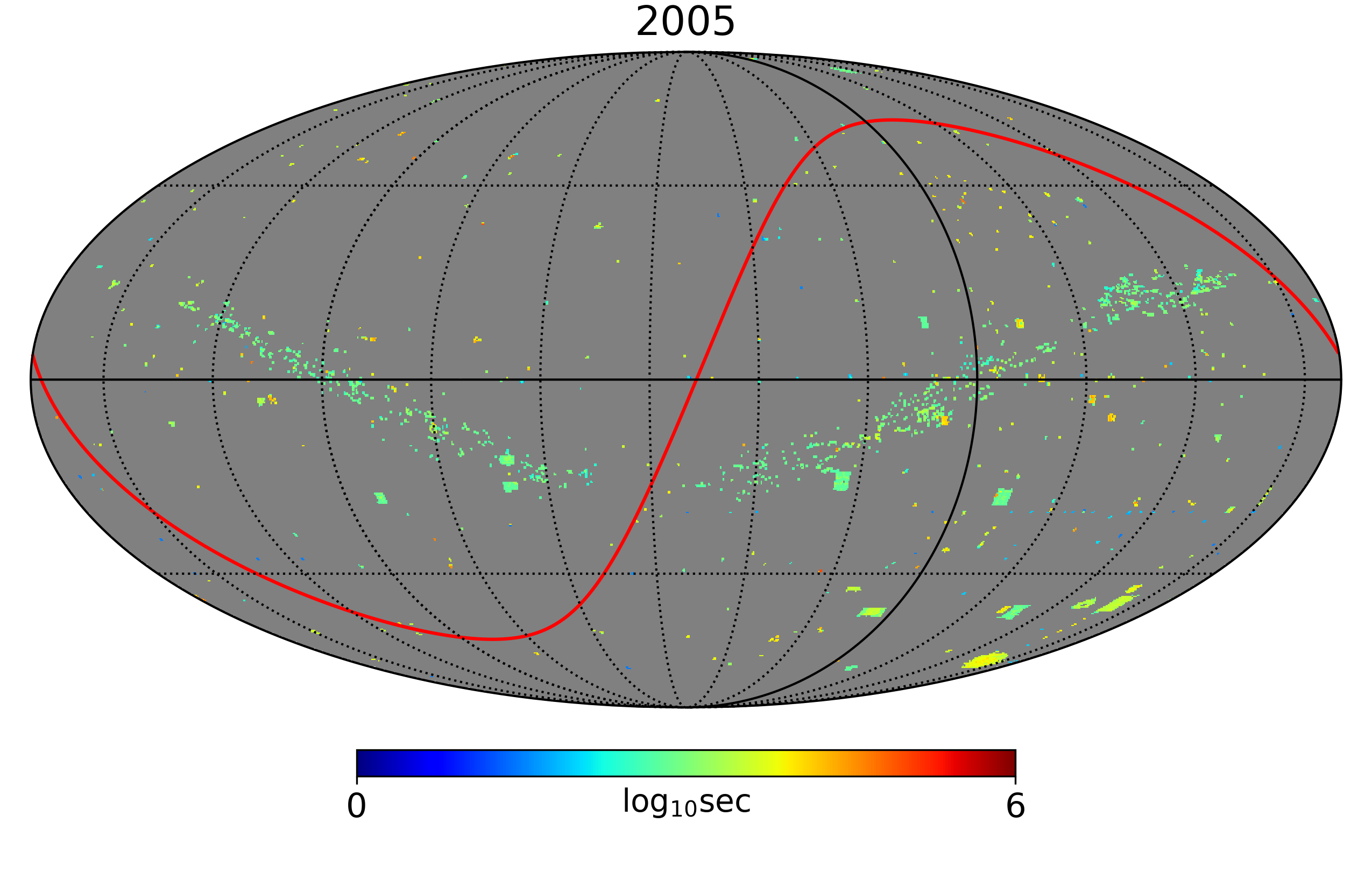}
    \end{subfigure} %
    \begin{subfigure}[b]{0.45\textwidth}
        \centering
        \includegraphics[width=\textwidth]{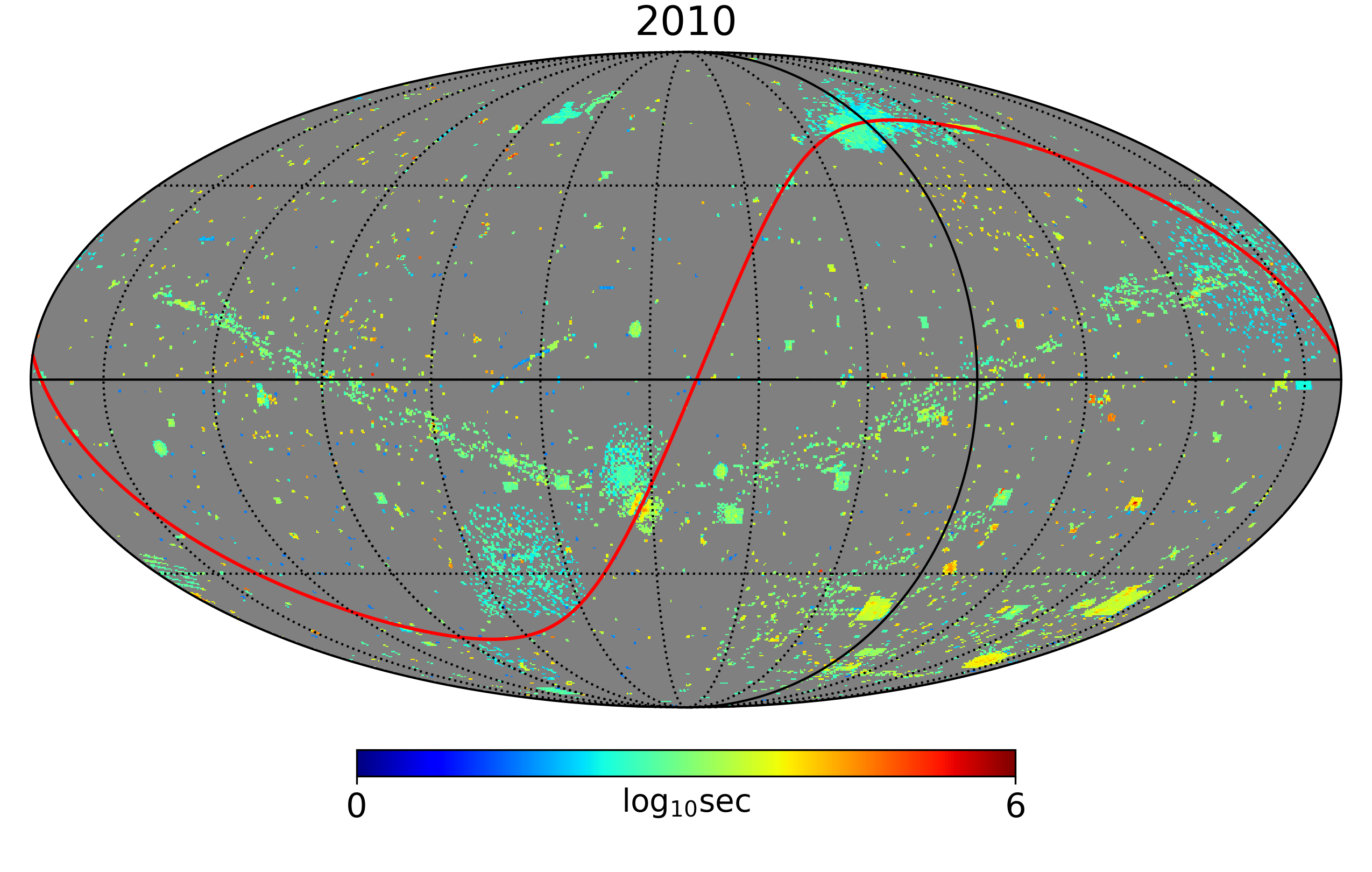}
    \end{subfigure} %
    \begin{subfigure}[b]{0.45\textwidth}
        \centering
        \includegraphics[width=\textwidth]{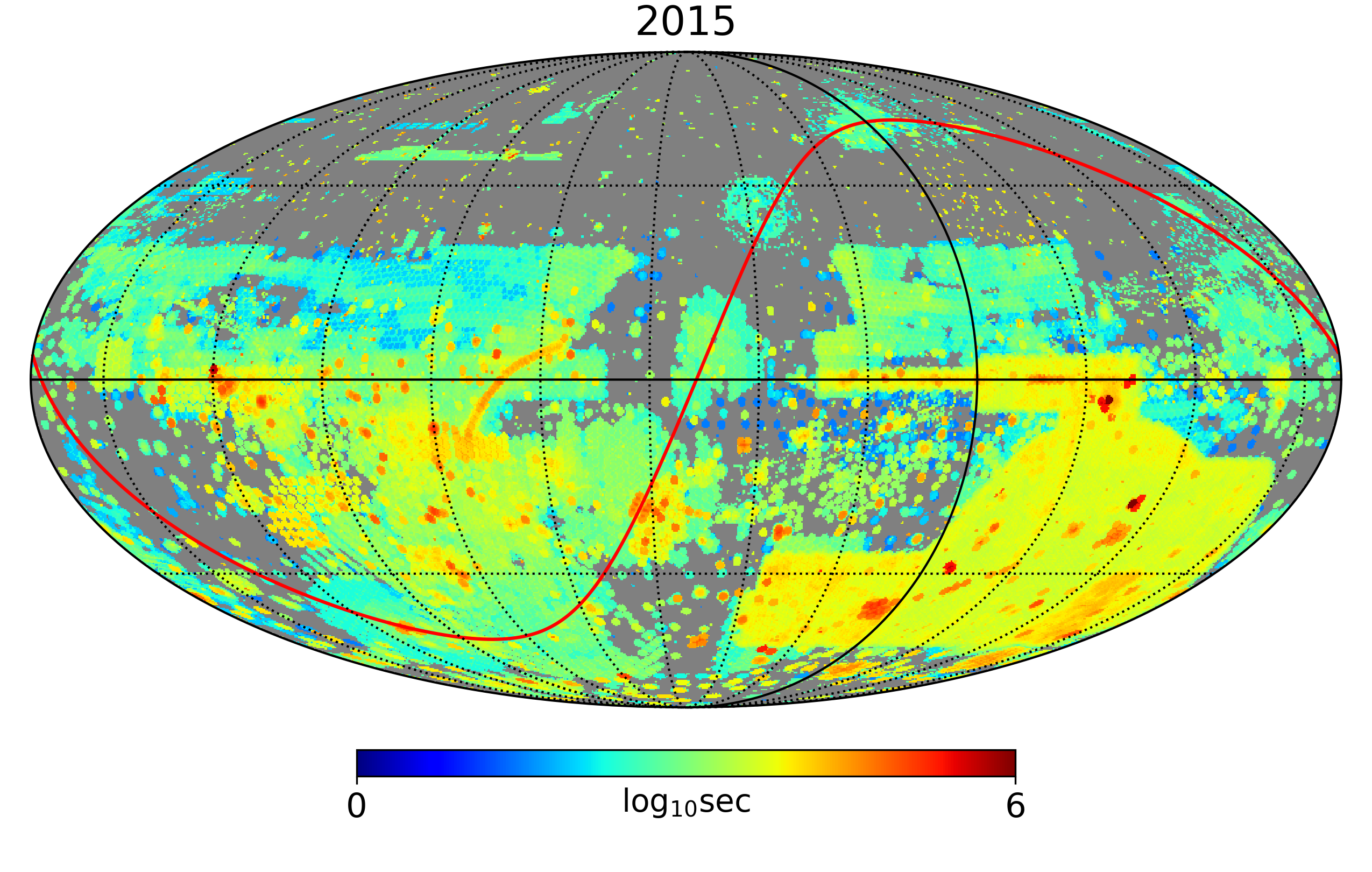}
    \end{subfigure}
    \begin{subfigure}[b]{0.45\textwidth}
        \centering
        \includegraphics[width=\textwidth]{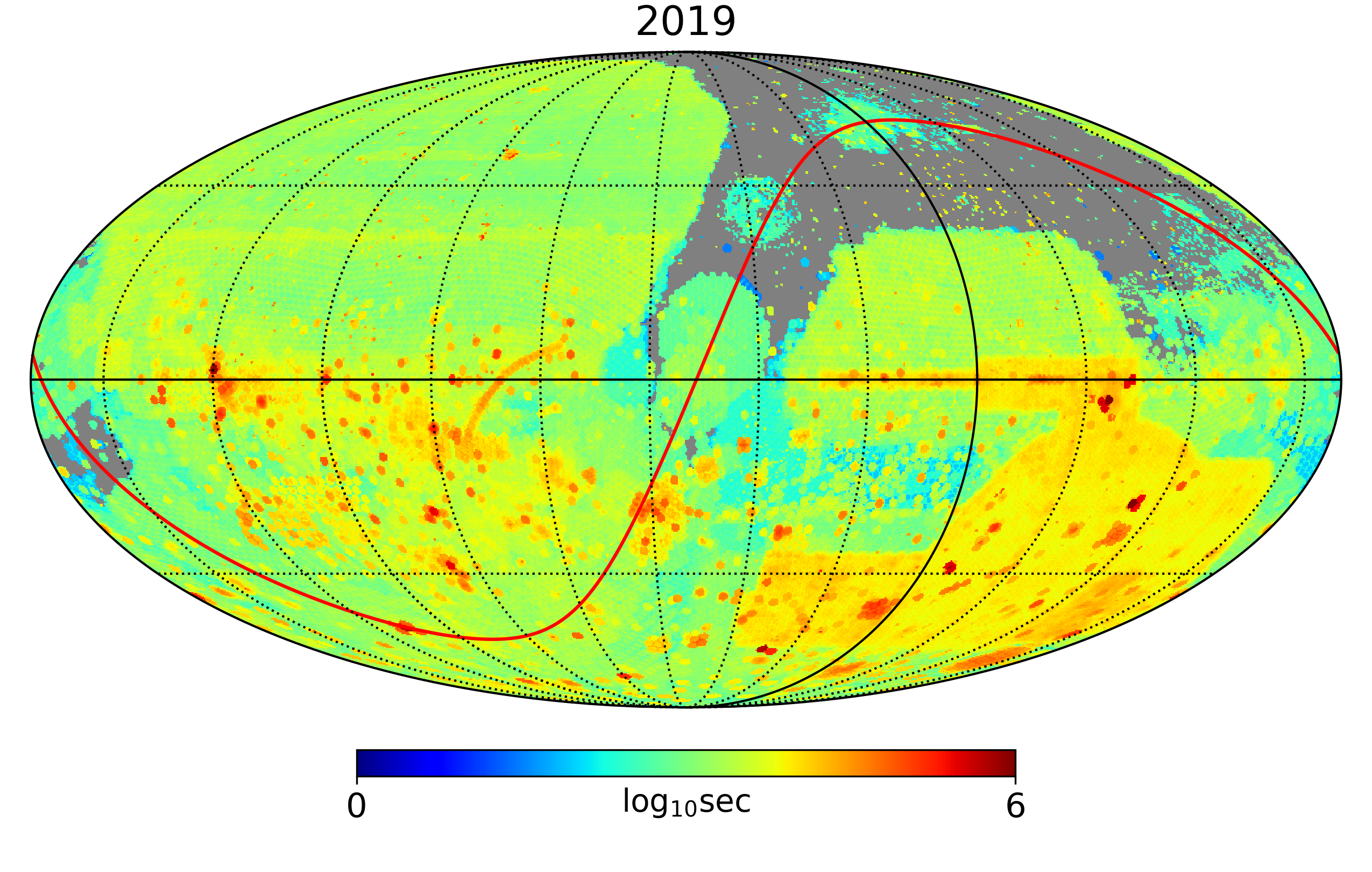}
    \end{subfigure}
\caption{Map of total exposure time for images from the Mosaic (1, 1.1, 2, 3) cameras and DECam in $\sim$5-year intervals, showing the growth in area and depth of imaging available in the NOAO Archive.}\label{expmap}
\end{figure}

NOAO has been a leader in providing the community with wide-field instruments on its telescopes for the past two decades. The Dark Energy Camera (DECam) and Dark Energy Survey (DES) currently provide the community with a cutting edge wide field camera, imaging data processing system, and data products, soon to be joined by the massively multiplexed Dark Energy Spectroscopic Instrument (DESI). With such capabilities, it is easy to predict continued growth in the size and complexity of NOAO-based surveys, and that more of the community will want to exploit them in their research.  As seen in Fig.~\ref{expmap}, over the last decade, imaging programs and large-scale surveys at NOAO have grown from covering a few percent of the sky to nearly its entirety, resulting in petabyte-scale data volume.  These imaging datasets have yielded very large catalogs; the NOAO Source Catalog \citep{2018AJ....156..131N}, which is an aperture photometry-based catalog of all of the public images, alone contains $\sim$3 billion objects and $\sim$34 billion rows of measurements \rob{and derived quantities.} 

The NOAO Data Lab project (\url{https://datalab.noao.edu}, \citealt{2014SPIE.9149E..1TF}) was launched in 2017
to enable efficient exploration and analysis of the large catalogs, and pixel- and spectral datasets derived from observations on NOAO-operated and other telescopes.  The Data Lab is a Science Platform that allows users to develop intuition through interaction with catalogs, images, and spectra, and to turn that intuition into discovery through automation of scientific analyses.  As we enter a decade that will feature the start of LSST operations and many other survey facilities (\rob{e.g. WFIRST, EUCLID, and MSE}), 
as well as a number of Science Platforms designed to exploit them (see Desai et al.\ 2019 APC white paper for a full list), the NOAO Data Lab will continue to serve an important role, as we describe in this paper.  Science Platforms will be especially crucial to maintaining a level playing field
as datasets grow to such size and complexity that they become difficult or impossible to handle for many individual researchers without supporting infrastructure.  {\bf We ask that Astro2020 recognize that Science Platform development and operations are a critical function of a U.S. National Observatory.  We also ask Astro2020 to acknowledge that Science Platforms will play a crucial role in realizing the scientific potential of the coming decade's surveys and facilities.}    

\section{Key Science Goals}
The goal of the NOAO Data Lab is to enable efficient archival use of massive survey data.  Several of the surveys being conducted on NOAO telescopes have been designed specifically with the goal of constraining the nature of Dark Energy (e.g. the Dark Energy Survey, \citealt{2018ApJS..239...18A}; Legacy Survey, \citealt{2019AJ....157..168D}; and the coming DESI Survey, \citealt{2016arXiv161100036D}).  Recent history has shown, however, that the impact of surveys often far exceed their original goals through archival use of the data.
For example, the 20-year SDSS survey has resulted in 7700 refereed papers containing SDSS data, the large majority from authors outside of the SDSS Collaboration (https://www.sdss.org/science/). Papers based on archival use of HST data outpace those based on GO programs \textcolor{black}{(376 {\it vs.} 341 in 2017; \citealt{HSTPUBSTAT})}.  In one year, Gaia DR2 has resulted in $\sim$300 refereed publications, $\sim$500 if arXiv preprints are included, thanks to the prompt public release of science-ready catalogs and databases.  
Along these lines, the Data Lab has been developing example archival use cases such as the following:


{\bf Catalog-based science cases:}
The dwarf companion galaxies of the Milky Way, Local Group, and Local Volume are critical probes both of the dark matter halos that are the seeds of forming galaxies and of the physical processes that shape their formation.  Their large mass-to-light ratios make them one of the most promising class of targets for probing the nature of dark matter (e.g.\ \citealt{2019BAAS...51c.207B}).

\begin{figure}
\centering
    \begin{subfigure}[b]{0.45\textwidth}
        \centering
        \includegraphics[width=\textwidth]{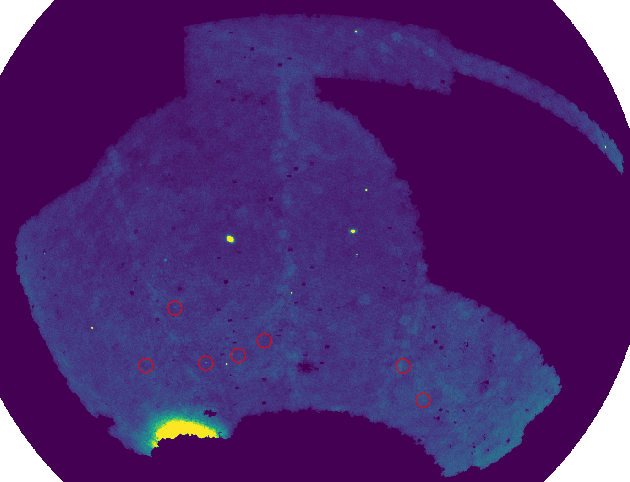}
    \end{subfigure} %
    \begin{subfigure}[b]{0.45\textwidth}
        \centering
        \includegraphics[width=\textwidth]{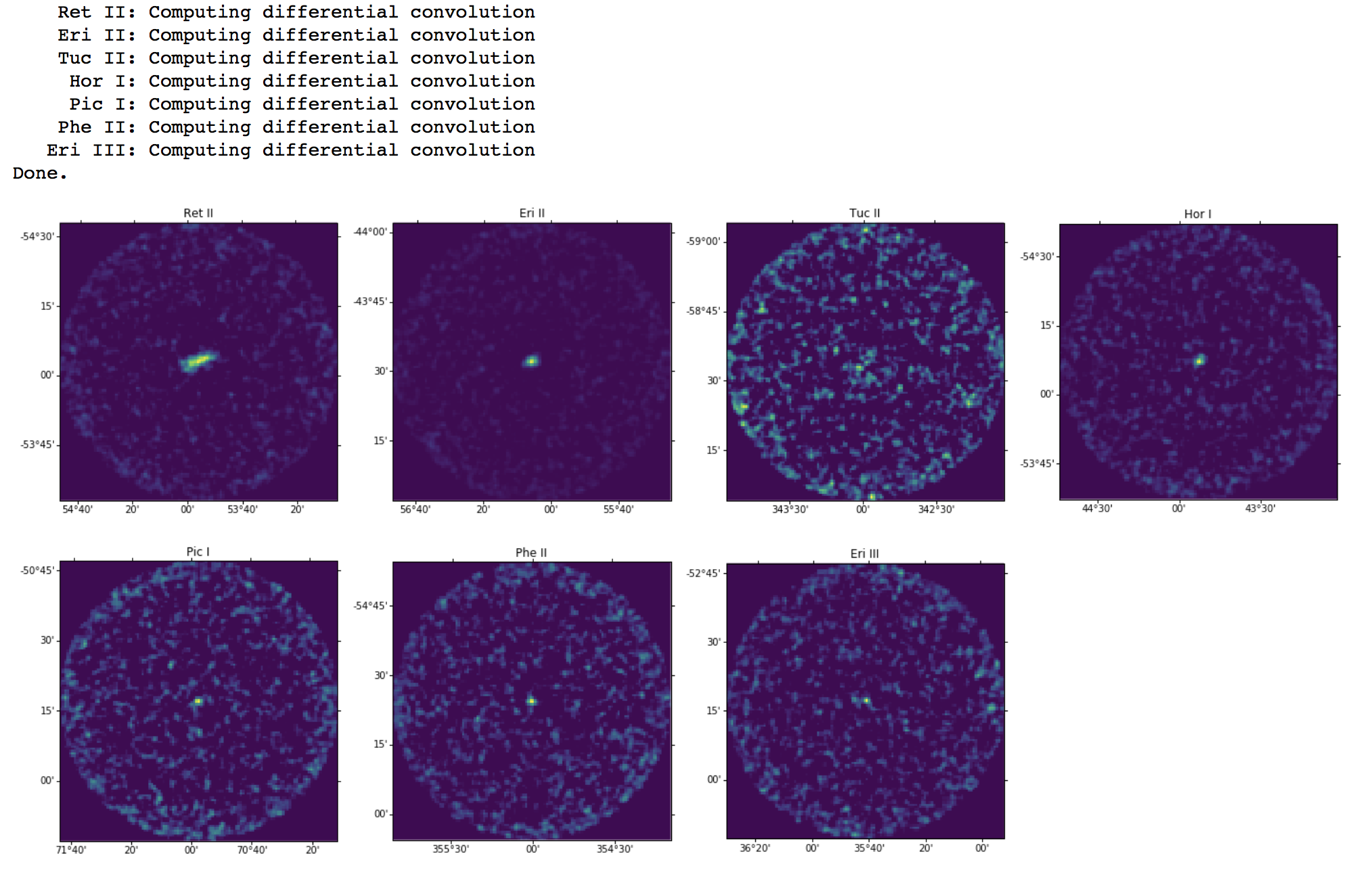}
    \end{subfigure} %
\caption{{\it Left:} Density map of point-like sources in DES with $-0.5<g-r<1.0$, isolating metal-poor stellar populations.  The outskirts of the LMC feature prominently at the southern edge of the map.  Red circles mark the locations of dwarfs discovered by \citet{2015ApJ...807...50B}, at the centers of which are faintly visible overdensities in the map. {\it Right:} Figure from the Data Lab science example notebook demonstrating the detection of seven dwarf galaxies in DES catalog data by K.\ Olsen and R.\ Nikutta (\url{https://github.com/noaodatalab/notebooks-latest/tree/master/03_ScienceExamples/DwarfGalaxies})}\label{dwarfs}
\end{figure}

Starting with SDSS and continuing with DECam and Gaia, catalog-based searches have yielded the discovery of $>$40 new dwarf satellites of the Milky Way, including several that may themselves be companions of the Magellanic Clouds \citep{2015ApJ...807...50B,2015ApJ...805..130K}.  The now well-established technique of discovery involves color-based selection and isochrone-based masking of catalog objects, followed by convolution with spatial filters and inspection of the original images for verification.  This technique results in a massive reduction of data, from hundreds of millions of photometric objects to hundreds of candidate dwarfs in the case of DES (see Fig.~\ref{dwarfs}).  An online Science Platform such as Data Lab eliminates the need to transfer Terabytes of catalog data, allows for the extraction of image cutouts around candidate objects, and makes the technique available for experimentation by anyone.  Many other science cases, such as the identification of stellar streams, star clusters, and galaxy clusters follow similar workflows, and would be well-supported by Data Lab and other Science Platforms.

\begin{figure}
\centering
\includegraphics[width=0.95\textwidth]{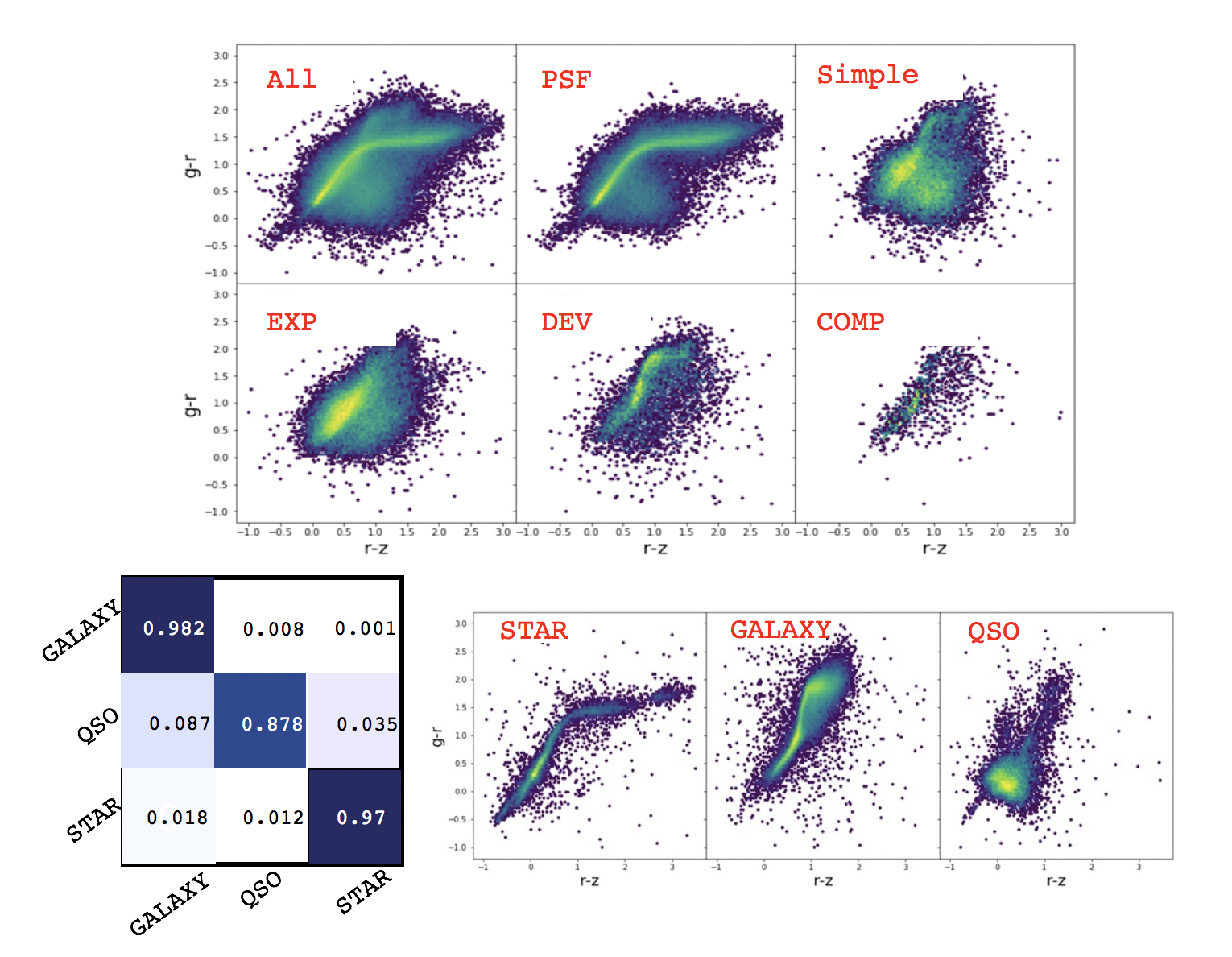}
\caption{Object classification in Legacy Survey DR7 as featured in the science example notebook by S.\ Juneau (\url{https://github.com/noaodatalab/notebooks-latest/blob/master/03_ScienceExamples/StarGalQSOSeparation/StarGalQsoLSDR7.ipynb}). {\it Top:} Color-color diagrams from a query retrieving 400,000 objects with S/N$>$3 in $grz$ from LS DR7, displayed according to the best-fitting model type as determined by the Tractor pipeline.  {\it Bottom right:} Color-color diagrams of LS DR3 objects cross-matched with SDSS DR13 spectroscopic classifications, courtesy of B.\ Abolfathi.  {\it Bottom left:} Confusion matrix from the Random Forest classifier trained using the cross-matched LS and SDSS objects, courtesy of J.-T. Schindler. }\label{classify}
\end{figure}

{\bf Object classification:}
Object classification is a critical step for almost all uses of archival survey data, including science cases that rely on star-galaxy separation, galaxy morphological classification, and QSO identification.  The Tractor \citep{2016ascl.soft04008L}, the pipeline used to reduce imaging from the Legacy Survey \citep{2019AJ....157..168D}, fits a sequence of morphological models to detected objects to aid in object classification.  These model fits, combined with photometric fluxes, form the basis for general object classification, as demonstrated in Fig.~\ref{classify}.  Using a subset of objects with spectroscopic identifications as a training set, the model fits, photometry, and other measurements may also be used to construct machine learning-based object classifiers.  Data Lab and other Science Platforms are ideal for developing such classifiers, as they aim to collect all of the relevant object and training data in one place (including diverse catalogs, images, and spectra), as well as provide tools for joining datasets (e.g. a cross-match service), a software development environment with the latest libraries, and sufficient hardware for performing the computations.  The availability of billions of photometric measurements and tens of millions of spectra from DESI, along with all of the Legacy Survey catalogs and images, will make Data Lab especially powerful for developing and training object classifiers.


{\bf Supporting time domain science:}
Astronomy in the time domain is a rapidly growing area of research (e.g. Graham et al.\ 2019), with time domain surveys playing central roles in several of the most important discoveries of the past two decades, such as that of Dark Energy (Reiss et al. 1998, Perlmutter et al. 1999), the existence of thousands of extra-solar planetary systems (e.g. Burke et al. 2014), and the detection of gravitational waves from merging black holes and neutron stars \citep{2017PhRvL.118v1101A,2017PhRvL.119p1101A}. The number of time domain surveys is growing accordingly, with the Zwicky Transient Facility \citep{2019PASP..131a8002B} logging $>10^5$ alerts of changes in brightness of variable and transient sources each night of operation, and the
10-year LSST survey poised to issue millions of such alerts per night.

Science Platforms have an important role to play for the coming decade of time domain science.  While the role of managing massive alert streams falls on time domain brokers such as \ANTARES\ (Matheson et al.\ 2019 APC white paper), which will allow researchers to narrow the alert streams to contain those events of highest interest, developing filters for use by brokers and testing them against databases of time series history is a natural function of Science Platforms, as Data Lab is currently doing for \ANTARES.  Once a set of alerts has been captured and classified by brokers, further 
investigation and analysis of the events would be strengthened by application of the full suite of datasets and tools available in a Science Platform such as Data Lab.  For the subset of alerts targeted for photometric or spectroscopic follow-up, Target and Observation Managers (TOMs, e.g.\ Street et al. 2018) are under development for managing the data acquisition.  Data Lab, which hosts SDSS spectroscopic data, will have tools for retrieving, visualizing, and analyzing spectra from DESI. Data Lab is under the same organizational umbrella as \ANTARES\ and follow-up facilities such as SOAR and Gemini, and would be an excellent platform for collecting and analyzing the follow-up data acquired through TOMs.


\section{Technical Overview}
Data Lab was designed to support several approaches to survey research, including: 1) Catalog-driven research, for which discoveries are made purely through sample selection from large catalogs; 2) Combined catalog and pixel research, for which images or spectra for a selection of catalog objects is important; 3) Custom workflows, for which complex or automated analyses of catalog objects and/or pixels is needed; and 4) Collaborative research, for which the work required to make a discovery depends on the
coordinated efforts of a team of people.  Data Lab provides the following services to support these approaches:

\begin{itemize}
    \item {\bf Data Discovery:} Users may browse available surveys and coverage through individual survey description pages, a catalog schema browser, and an interactive sky viewer.  As of this writing, the Data Lab hosts catalogs and images from 17 distinct surveys covering multiple data releases.

    \item {\bf Database queries:} Data Lab's catalog holdings are $\sim$50 TB in volume and contain $\sim$150 billion rows of data.  These are searchable through a Python-based client that accepts synchronous and asynchronous queries written in SQL or ADQL, through a web-based ADQL query form, and through clients such as TOPCAT \citep{2005ASPC..347...29T}.

    \item {\bf Data storage:} Data Lab provides personal database storage (myDB) for user tables, as well as user file storage backed by a large distributed storage volume.  Each user is given a public folder for sharing contents with collaborators.

    \item {\bf File services:} Data Lab provides read-only access to $\sim$600 TB of file-based data products, including survey-specific image products, spectra, and tabular data.

    \item {\bf Catalog cross-matches:} By uploading catalogs to myDB tables, users may use the Data Lab Jupyter notebook server or a web interface to perform fast cross-matches of personal catalogs with hosted catalogs.

    \item {\bf Image cutouts:} Image cutouts are available through Data Lab from the full petabyte-scale NOAO Science Archive, survey-specific collections, as well \rob{as} collections of raw, reduced, and coadded images.  Image cutouts may be performed at command level through Python or through a web interface.

    \item {\bf Data analysis:} Data Lab provides authenticated access to a Jupyter notebook server, to which
    all services are exposed through APIs and Python clients.

    \item {\bf Documentation and support:} Documentation is provided in the form of Jupyter notebooks that provide tutorials, detailed instructions on specific services, and complete science examples; an online user guide; API documentation; and an online helpdesk.  Users are encouraged to contribute their own example notebooks through the public GitHub notebook repository.

\end{itemize}

\section{Technology Drivers}
The Data Lab architecture was designed to be modular and to make use of as much existing development and technology as possible, and will continue to do so for its future development.  Services are exposed through Application Programming Interfaces (APIs) and programmatic clients, with web-based interfaces developed as wrappers around these APIs.

Data Lab makes extensive use of International Virtual Observatory Alliance (IVOA; \url{http://www.ivoa.net/}) protocols and implementations developed elsewhere in its services.  In particular, Data Lab uses:

\begin{itemize}
    \item Table Access Protocol (TAP), based on the implementation by CADC
    \item Simple Image Access (SIA) protocol
    \item VOSpace based on the implementation by CADC (\url{https://www.canfar.net/en/docs/storage/})
    \item Universal Worker Service (UWS) standard
    \item DALServer based on the implementation by NRAO (\url{https://github.com/TomMcGlynn/usvirtualobservatory/tree/master/usvao/prototype/dalserver})
    \item PyVO Python library (\url{https://pyvo.readthedocs.io/en/latest/})
\end{itemize}

Data Lab also uses several Open Source software projects as part of its services, including:

\begin{itemize}
    \item PostgreSQL (\url{https://www.postgresql.org}) as its large catalog database solution
    \item Q3C \citep{2019ascl.soft05008K} for spatial indexing, cone searches, and cross-matching
    \item Project Jupyter (\url{https://jupyter.org}) as its analysis platform
    \item AstroPy \citep{astropy:2013,astropy:2018} and affiliated packages
\end{itemize}


Data Lab is actively developing the use of Docker (\url{https://www.docker.com}) Containers and Container orchestration through Kubernetes (\url{https://kubernetes.io}) for service management and deployment.  While we have not yet found commercial Cloud deployment to be a cost effective solution for Data Lab, exploration of Cloud deployment  continues to be part of the technological development roadmap.  

\section{Organization, Partnerships, and Current Status}
Data Lab is a core project of the NOAO/NCOA Community Science and Data Center (CSDC), as described in the Bolton et al.\ (2019) APC white paper.  Data Lab shares development goals with the \ANTARES\ project \citep{2014htu..conf..145M} and with Data Management Operations, which develops and maintains the NOAO Science Archive.  The Data Lab mission and future development align well with two of the NSF's Big Ideas\footnote{https://www.nsf.gov/about/congress/reports/nsf\_big\_ideas.pdf}, namely {\it Harnessing Data for 21st Century Science and Engineering} and as part of a network to contribute to {\it Windows on the Universe: The Era of Multi-messenger Astrophysics}.

Data Lab is working with the DESI Project to host its public imaging catalogs and to develop spectroscopic visualization and analysis tools compatible with DESI spectroscopic data.

As part of NCOA, Data Lab will fall under the same organizational umbrella as LSST and the Gemini Observatory.  As part of its technology development roadmap, Data Lab aims to ensure future compatibility with the LSST Science Platform (\url{https://ldm-542.lsst.io}).  As part of a Science Platform Network (Desai et al.\ 2019 APC white paper), \rob{Data Lab} plans to develop its services in a cooperative manner with other Science Platform efforts.

\begin{figure}
\centering
    \begin{subfigure}[b]{0.45\textwidth}
        \centering
        \includegraphics[width=\textwidth]{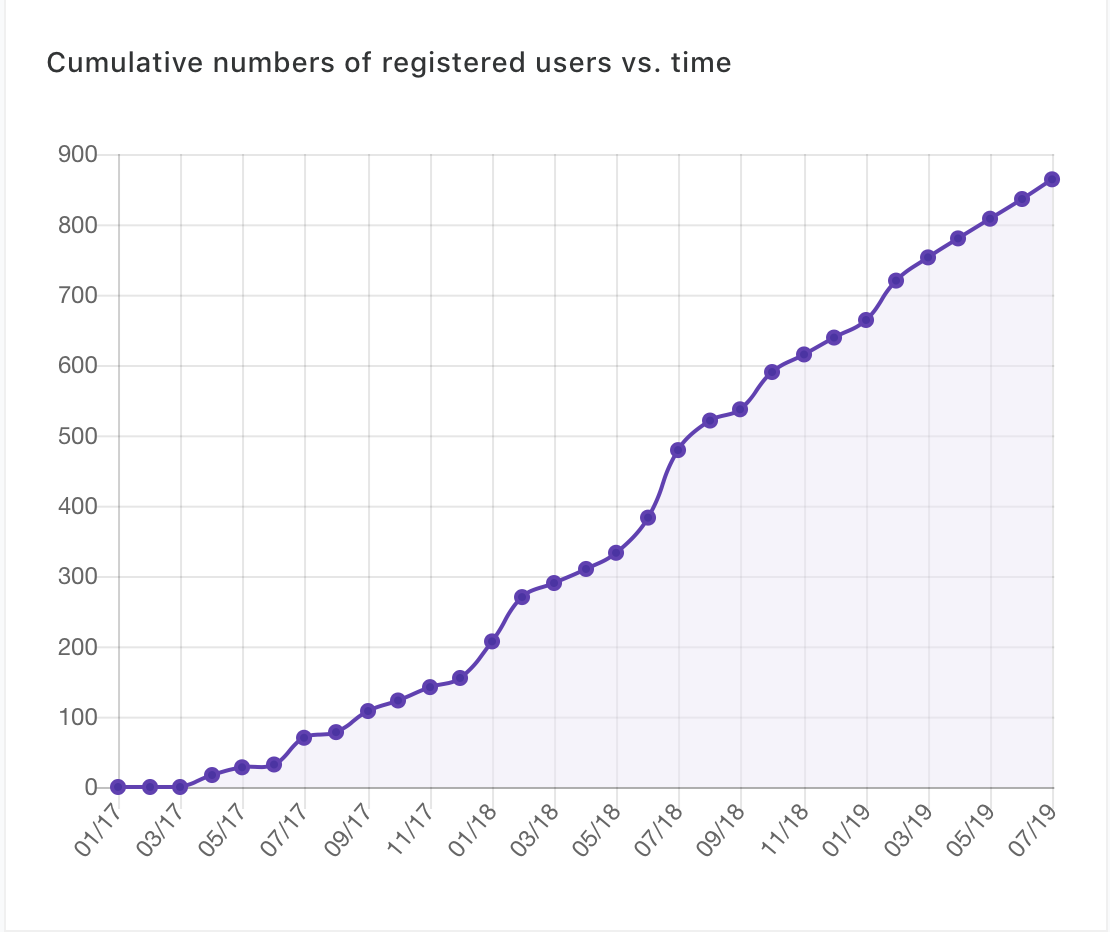}
    \end{subfigure} %
    \begin{subfigure}[b]{0.45\textwidth}
        \centering
        \includegraphics[width=\textwidth]{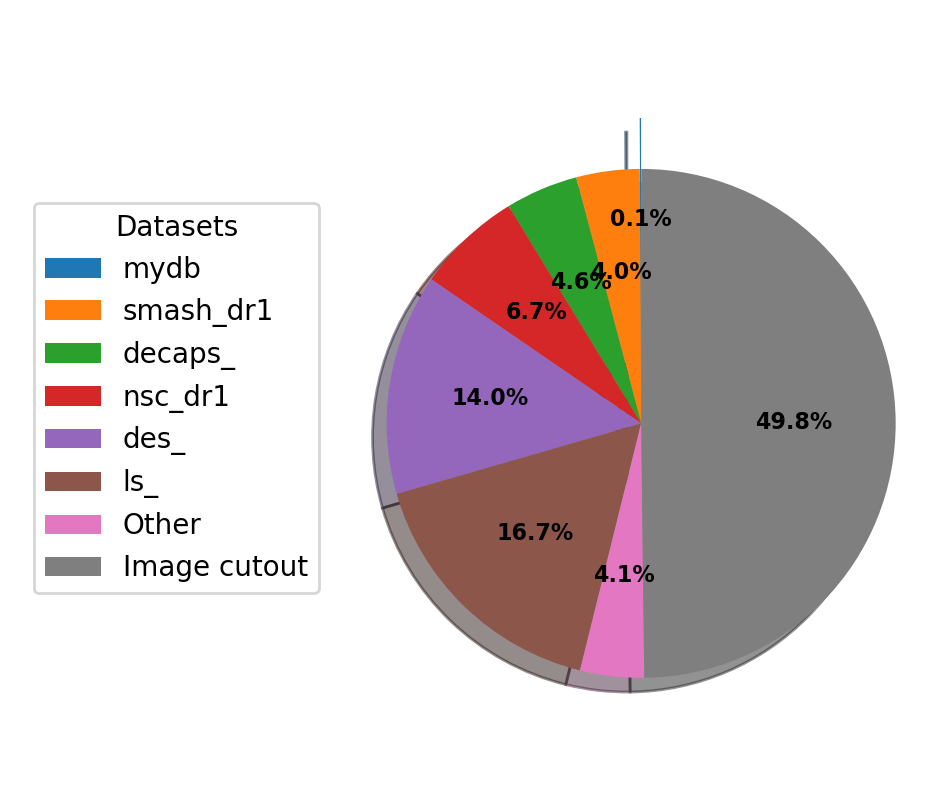}
    \end{subfigure} %
\caption{{\it Left:} Growth of registered Data Lab users since its June 2017 public release.  As of this writing, Data Lab has 864 registered users.  {\it Right:} Breakdown of which Data Lab datasets were queried most frequently by users over a recent 8-month period.  Roughly 50\% of the $\sim$6000 daily queries were for image cutouts, while DECam- and Mosaic-based catalogs accounted for $\sim$45\% of the queries.  Queries against myDB represent interactive catalog uploads and cross-matches by users; their small relative number indicates that the vast majority of queries are automated scripted queries.}\label{dlusage}
\end{figure}

As of this writing, Data Lab has 864 registered users, with the number of new users continuing to grow linearly with time (Fig.~\ref{dlusage}).  Over a recent eight-month period, Data Lab received, on average, $>$6000 queries of its datasets per day, with the vast majority of these likely being scripted, automated queries (Fig.~\ref{dlusage}).  Of those queries, $\sim$50\% were image cutout requests, while the majority of the remainder were queries against DECam- and Mosaic-based catalogs.
Over roughly that same period, Data Lab received $\sim$5 unique interactive visitors per day, as evidenced by the rate of interactive catalog cross-match queries and Jupyter notebook server usage.  These statistics indicate that Data Lab is meeting its overall goal of supporting exploration of the data produced on NOAO telescopes.

\section{Schedule}
The major Data Lab development schedule has been as follows:
\begin{itemize}
    \item August 2014: Project Kick-off
    \item March 2015: Conceptual Design Review by external panel
    \item June 2016: First public demo
    \item August 2016: Interim Review by external panel
    \item June 2017: Data Lab v1.0 release
    \item June 2018: Data Lab v2.0 release
\end{itemize}
Data Lab is now on a faster incremental release schedule, with current release version 2.17.1, aimed at deploying small improvements in a continual fashion.  

\section{Cost Estimates}
The cost of developing and operating Data Lab is $\sim$7.5 FTE per year, with total cost of $\sim$\$1 million per year.  Most of the cost is personnel, while equipment and travel account for $\sim$\$115,000 per year.




\let\oldbibliography\thebibliography
\renewcommand{\thebibliography}[1]{\oldbibliography{#1}
\setlength{\itemsep}{0pt}} 
\bibliography{DataLab}

\begin{thebibliography}{}
\expandafter\ifx\csname natexlab\endcsname\relax\def\natexlab#1{#1}\fi

\bibitem[{{Abbott} {et~al.}(2017{\natexlab{a}}){Abbott}, {Abbott}, {Abbott},
  {Acernese}, {Ackley}, {Adams}, {Adams}, {Addesso}, {Adhikari}, \&
  {Adya}}]{2017PhRvL.118v1101A}
{Abbott}, B.~P., {Abbott}, R., {Abbott}, T.~D., {et~al.} 2017{\natexlab{a}},
  \prl, 118, 221101

\bibitem[{{Abbott} {et~al.}(2017{\natexlab{b}}){Abbott}, {Abbott}, {Abbott},
  {Acernese}, {Ackley}, {Adams}, {Adams}, {Addesso}, {Adhikari}, \&
  {Adya}}]{2017PhRvL.119p1101A}
---. 2017{\natexlab{b}}, \prl, 119, 161101

\bibitem[{{Abbott} {et~al.}(2018){Abbott}, {Abdalla}, {Allam}, {Amara},
  {Annis}, {Asorey}, {Avila}, {Ballester}, {Banerji}, \&
  {Barkhouse}}]{2018ApJS..239...18A}
{Abbott}, T.~M.~C., {Abdalla}, F.~B., {Allam}, S., {et~al.} 2018, \apjs, 239,
  18

\bibitem[{{Astropy Collaboration} {et~al.}(2013){Astropy Collaboration},
  {Robitaille}, {Tollerud}, {Greenfield}, {Droettboom}, {Bray}, {Aldcroft},
  {Davis}, {Ginsburg}, {Price-Whelan}, {Kerzendorf}, {Conley}, {Crighton},
  {Barbary}, {Muna}, {Ferguson}, {Grollier}, {Parikh}, {Nair}, {Unther},
  {Deil}, {Woillez}, {Conseil}, {Kramer}, {Turner}, {Singer}, {Fox}, {Weaver},
  {Zabalza}, {Edwards}, {Azalee Bostroem}, {Burke}, {Casey}, {Crawford},
  {Dencheva}, {Ely}, {Jenness}, {Labrie}, {Lim}, {Pierfederici}, {Pontzen},
  {Ptak}, {Refsdal}, {Servillat}, \& {Streicher}}]{astropy:2013}
{Astropy Collaboration}, {Robitaille}, T.~P., {Tollerud}, E.~J., {et~al.} 2013,
  \aap, 558, A33

\bibitem[{{Bechtol} {et~al.}(2015){Bechtol}, {Drlica-Wagner}, {Balbinot},
  {Pieres}, {Simon}, {Yanny}, {Santiago}, {Wechsler}, {Frieman}, \&
  {Walker}}]{2015ApJ...807...50B}
{Bechtol}, K., {Drlica-Wagner}, A., {Balbinot}, E., {et~al.} 2015, \apj, 807,
  50

\bibitem[{{Bechtol} {et~al.}(2019){Bechtol}, {Drlica-Wagner}, {Abazajian},
  {Abidi}, {Adhikari}, {Ali-Ha{\i}{\ensuremath{\ddot{}}}moud}, {Annis},
  {Ansarinejad}, {Armstrong}, \& {Asorey}}]{2019BAAS...51c.207B}
{Bechtol}, K., {Drlica-Wagner}, A., {Abazajian}, K.~N., {et~al.} 2019, in
  \baas, Vol.~51, 207

\bibitem[{{Bellm} {et~al.}(2019){Bellm}, {Kulkarni}, {Graham}, {Dekany},
  {Smith}, {Riddle}, {Masci}, {Helou}, {Prince}, \&
  {Adams}}]{2019PASP..131a8002B}
{Bellm}, E.~C., {Kulkarni}, S.~R., {Graham}, M.~J., {et~al.} 2019, \pasp, 131,
  018002

\bibitem[{{DESI Collaboration} {et~al.}(2016){DESI Collaboration}, {Aghamousa},
  {Aguilar}, {Ahlen}, {Alam}, {Allen}, {Allende Prieto}, {Annis}, {Bailey}, \&
  {Balland}}]{2016arXiv161100036D}
{DESI Collaboration}, {Aghamousa}, A., {Aguilar}, J., {et~al.} 2016, arXiv
  e-prints, arXiv:1611.00036

\bibitem[{{Dey} {et~al.}(2019){Dey}, {Schlegel}, {Lang}, {Blum}, {Burleigh},
  {Fan}, {Findlay}, {Finkbeiner}, {Herrera}, \& {Juneau}}]{2019AJ....157..168D}
{Dey}, A., {Schlegel}, D.~J., {Lang}, D., {et~al.} 2019, \aj, 157, 168

\bibitem[{{Fitzpatrick} {et~al.}(2014){Fitzpatrick}, {Olsen}, {Economou},
  {Stobie}, {Beers}, {Dickinson}, {Norris}, {Saha}, {Seaman}, \&
  {Silva}}]{2014SPIE.9149E..1TF}
{Fitzpatrick}, M.~J., {Olsen}, K., {Economou}, F., {et~al.} 2014, in Society of
  Photo-Optical Instrumentation Engineers (SPIE) Conference Series, Vol. 9149,
  \procspie, 91491T

\bibitem[{{HST Library Staff}(2018)}]{HSTPUBSTAT}
{HST Library Staff}. 2018, HST Publication Statistics,
  \url{https://archive.stsci.edu/hst/bibliography/pubstat.html}

\bibitem[{{Koposov} \& {Bartunov}(2019)}]{2019ascl.soft05008K}
{Koposov}, S., \& {Bartunov}, O. 2019, {Q3C: A PostgreSQL package for spatial
  queries and cross-matches of large astronomical catalogs}, ascl:1905.008

\bibitem[{{Koposov} {et~al.}(2015){Koposov}, {Belokurov}, {Torrealba}, \&
  {Evans}}]{2015ApJ...805..130K}
{Koposov}, S.~E., {Belokurov}, V., {Torrealba}, G., \& {Evans}, N.~W. 2015,
  \apj, 805, 130

\bibitem[{{Lang} {et~al.}(2016){Lang}, {Hogg}, \&
  {Mykytyn}}]{2016ascl.soft04008L}
{Lang}, D., {Hogg}, D.~W., \& {Mykytyn}, D. 2016, {The Tractor: Probabilistic
  astronomical source detection and measurement}, ascl:1604.008

\bibitem[{{Matheson} {et~al.}(2014){Matheson}, {Saha}, {Snodgrass}, \&
  {Kececioglu}}]{2014htu..conf..145M}
{Matheson}, T., {Saha}, A., {Snodgrass}, R., \& {Kececioglu}, J. 2014, in The
  Third Hot-wiring the Transient Universe Workshop, ed. P.~R. {Wozniak}, M.~J.
  {Graham}, A.~A. {Mahabal}, \& R.~{Seaman}, 145--150

\bibitem[{{Nidever} {et~al.}(2018){Nidever}, {Dey}, {Olsen}, {Ridgway},
  {Nikutta}, {Juneau}, {Fitzpatrick}, {Scott}, \&
  {Valdes}}]{2018AJ....156..131N}
{Nidever}, D.~L., {Dey}, A., {Olsen}, K., {et~al.} 2018, \aj, 156, 131

\bibitem[{{Price-Whelan} {et~al.}(2018){Price-Whelan}, {Sip{\H{o}}cz},
  {G{\"u}nther}, {Lim}, {Crawford}, {Conseil}, {Shupe}, {Craig}, {Dencheva},
  {Ginsburg}, {VanderPlas}, {Bradley}, {P{\'e}rez-Su{\'a}rez}, {de Val-Borro},
  {Paper Contributors}, {Aldcroft}, {Cruz}, {Robitaille}, {Tollerud},
  {Coordination Committee}, {Ardelean}, {Babej}, {Bach}, {Bachetti}, {Bakanov},
  {Bamford}, {Barentsen}, {Barmby}, {Baumbach}, {Berry}, {Biscani}, {Boquien},
  {Bostroem}, {Bouma}, {Brammer}, {Bray}, {Breytenbach}, {Buddelmeijer},
  {Burke}, {Calderone}, {Cano Rodr{\'\i}guez}, {Cara}, {Cardoso}, {Cheedella},
  {Copin}, {Corrales}, {Crichton}, {D{\textquoteright}Avella}, {Deil},
  {Depagne}, {Dietrich}, {Donath}, {Droettboom}, {Earl}, {Erben}, {Fabbro},
  {Ferreira}, {Finethy}, {Fox}, {Garrison}, {Gibbons}, {Goldstein}, {Gommers},
  {Greco}, {Greenfield}, {Groener}, {Grollier}, {Hagen}, {Hirst}, {Homeier},
  {Horton}, {Hosseinzadeh}, {Hu}, {Hunkeler}, {Ivezi{\'c}}, {Jain}, {Jenness},
  {Kanarek}, {Kendrew}, {Kern}, {Kerzendorf}, {Khvalko}, {King}, {Kirkby},
  {Kulkarni}, {Kumar}, {Lee}, {Lenz}, {Littlefair}, {Ma}, {Macleod},
  {Mastropietro}, {McCully}, {Montagnac}, {Morris}, {Mueller}, {Mumford},
  {Muna}, {Murphy}, {Nelson}, {Nguyen}, {Ninan}, {N{\"o}the}, {Ogaz}, {Oh},
  {Parejko}, {Parley}, {Pascual}, {Patil}, {Patil}, {Plunkett}, {Prochaska},
  {Rastogi}, {Reddy Janga}, {Sabater}, {Sakurikar}, {Seifert}, {Sherbert},
  {Sherwood-Taylor}, {Shih}, {Sick}, {Silbiger}, {Singanamalla}, {Singer},
  {Sladen}, {Sooley}, {Sornarajah}, {Streicher}, {Teuben}, {Thomas},
  {Tremblay}, {Turner}, {Terr{\'o}n}, {van Kerkwijk}, {de la Vega}, {Watkins},
  {Weaver}, {Whitmore}, {Woillez}, {Zabalza}, \& {Contributors}}]{astropy:2018}
{Price-Whelan}, A.~M., {Sip{\H{o}}cz}, B.~M., {G{\"u}nther}, H.~M., {et~al.}
  2018, \aj, 156, 123

\bibitem[{{Taylor}(2005)}]{2005ASPC..347...29T}
{Taylor}, M.~B. 2005, in Astronomical Society of the Pacific Conference Series,
  Vol. 347, Astronomical Data Analysis Software and Systems XIV, ed.
  P.~{Shopbell}, M.~{Britton}, \& R.~{Ebert}, 29

\end{thebibliography}

\end{document}